\documentclass{article}
\usepackage{fullpage, amsfonts, amsmath, hyperref, xcolor}
\def\bbC{\mathbb{C}}
\def\bbR{\mathbb{R}}
\def\bpm#1\epm{\begin{pmatrix}#1\end{pmatrix}}
\DeclareMathOperator{\poly}{poly}

\begin{document}
\title{Why now is the right time to study quantum computing}
\author{Aram W. Harrow\thanks{Center for Theoretical Physics, MIT}}
\maketitle

Quantum computing is a good way to justify difficult physics experiments.  But until quantum computers are built, do computer scientists need to know anything about quantum information?  In fact, quantum computing is not merely a recipe for new computing devices, but a new way of looking at the world that has been astonishingly intellectually productive.  In this article, I'll talk about where quantum computing came from, what it is, and what we can learn from it.

\section{The development of quantum computing}
In previous centuries, our deterministic view of the world led us to imagine it as a giant clockwork mechanism.  But as computers have become ubiquitous, they have also taken over our metaphors and changed the way we think about science, math and even society.  Not only do we use computers to solve problems, we also think in ways that are informed by building, programming and using computers.

For example, phenomena as diverse as DNA, language and cognition are now thought of as mechanisms for information transmission that have evolved to optimize for both compression and error correction.  Game theory and economics have begun to incorporate notions of computational efficiency, realizing, as Kamil Jain famously said of Nash equilibria, ``If your laptop can't find it, then neither can the market.'' Computer science has also reshaped the goals of these fields.  Mathematics increasingly concerns itself with matters of efficiency, and has been rapidly growing in the computer-related fields of information theory, graph theory and statistics.  The  P-vs-NP question is the newest of the Clay Millenium problems, and if resolved would shed light on the oldest puzzle in mathematics: what makes it hard to find proofs?

In hindsight, the computational view seems natural.  But when computers first appeared, even those few who predicted their commercial success could not have foreseen the intellectual revolution they would cause.  For example, entropy (which is central to compression and error correction) could easily have been invented in the time of Gauss or even by medieval Arabs or ancient Greeks.  But in practice the impetus to develop it came first in the $19^{\text{th}}$ century, when thermodynamics was developed to understand steam engines, and again in the $20^{\text{th}}$ century, when Bell Labs employed Claude Shannon under a wartime contract to study cryptography.   This is not unique to computer science; Einstein's work as a patent clerk helped him devise relativity, as clock synchronization on railway networks was an important engineering problem, and metaphors of clocks on trains provided grist for his famous thought experiments.  In general, science often follows technology because inventions give us new ways to think about the world and new phenomena in need of explanation.

The story of quantum computing is similar.  Quantum mechanics was invented in the first few decades of the $20^{\text{th}}$ century, and its modern form was mostly known by 1930. But the idea that quantum mechanics might give computational advantages did not arise until physicists tried to simulate quantum mechanics on computers.  Doing so, they quickly ran into a practical problem: while a single system (e.g., the polarization of a photon) might be describable by only two complex numbers (e.g., the amplitudes of the horizontal and vertical components of the polarization), a collection of $n$ such systems requires not $2n$ but $2^n$ complex numbers to fully describe, even though measurement can extract only $n$ bits.  Physicists coped by developing closed-form solutions and physically motivated approximations that could handle an increasingly large number of cases of interest (more on this later).

The exponentially large state spaces of quantum mechanics should have been a clue that nature contains vastly more computational resources than we ever imagined.  Instead, it and other strange features of quantum mechanics were seen mostly as limitations and quirks of the theory.  For example, Heisenberg's uncertainty principle was often thought of as a restriction on measurements.  Phenomena such as entanglement were considered part of ``quantum foundations'' or the philosophy of quantum mechanics, but were not considered operationally relevant until quantum computing and quantum cryptography were independently developed in the 1970s and 80s.

Quantum computing (or more precisely, the idea of quantum advantage in computing) came from Richard Feynman's 1982 suggestion that since conventional computers appeared to require exponential overhead to simulate quantum mechanics, perhaps a quantum-mechanical computer could perform the task more efficiently.  The model was formalized in 1985 by David Deutsch, who also showed, surprisingly, that a quantum-mechanical computer could be faster than a conventional computer for a problem (computing the XOR of two bits) that on its face had nothing to do with quantum mechanics.  A series of stronger speedups followed, albeit for contrived problems, until Peter Shor's 1994 quantum algorithm for factoring integers in polynomial time.

Much earlier (1970), then grad student Stephen Wiesner proposed using Heisenberg's restrictions on measurements to prevent an adversary from learning a secret message.  Thus quantum cryptography was born, although Wiesner's paper was almost universally rejected from journals, and its ideas were largely unknown until Charles Bennett and Gilles Brassard published a scheme for quantum key distribution in 1984 (and even this wasn't taken seriously until they implemented their proposal in 1991).

The key conceptual shift enabling both quantum computing and quantum cryptography was to start thinking about the effects of quantum mechanics on information in operational terms instead of as a source of limitations, curiosities and paradoxes.  Once this shift was made, the technical aspects of quantum information were much simpler than earlier developments in quantum mechanics, such as the unification in the 1950s of quantum mechanics with special relativity.

\section{Computing with qubits}
One of the basic contributions of the computational approach to quantum mechanics is the idea of a qubit.  In general, a quantum system with d perfectly distinguishable states can be described as a unit vector in $\bbC^d$.  The simplest interesting case is when $d=2$, and the resulting system is called a qubit.  Measuring the vector $x = \bpm x_0 \\ x_1 \epm$ yields outcome 0 with probability $|x_0|^2$  and outcome 1 with probability $|x_1|^2$; hence the requirement that $x$ be a unit vector.  Any linear dynamics are possible, as long as they preserve the norm of the state.  In other words, evolution is given by mapping $x$ to $Ux$, where $U$ is a unitary matrix, meaning that it always preserves length.  Mathematically, this is equivalent to the equation $U^\dag U = I$ , where $(U^\dag)_{ij} = \bar U_{ji}$.  The beauty of this explanation is that it is completely independent of the underlying system, be it a photon's polarization, the energy level of an electron bound to an atom, the spin of a nucleus or the direction of a loop of superconducting current.  In this way, a qubit is a {\em device-independent} way of describing quantum information, just as bits enable us to reason about classical information without needing to know whether it is encoded in RAM, a hard drive, or an abacus.

Single qubits are interesting for physics experiments, but for computational purposes, we are more interested in what happens when we have $n$ qubits.  In this case, the state $x$ is a unit vector in $\bbC^{2^n}$, with entries labeled by $n$-bit strings.  Most $n$-qubit states are entangled, meaning that their amplitudes are in some way correlated across the $n$ bits.  Dynamics are now given by  unitary matrices, which are built out of a series of two-qubit gates.  To represent a gate acting only on qubits, say, 3 and 7, we consider a unitary matrix $U$ in which $U_{ij}$ is nonzero only if the $n$-bit strings $i$ and $j$ are equal everywhere except on bits 3 and 7.  Moreover, the value of $U_{ij}$ should depend only the four bits $i_3$, $j_3$, $i_7$ and $j_7$.

This linear-algebra formalism may seem abstract, but can also be used to describe classical deterministic and randomized computations.  An $n$-bit string can be represented, albeit somewhat wastefully, as a vector of length $2^n$ with one entry equal to 1 and the rest equal to 0.  Dynamics can be represented by mapping $x$ to $Mx$ where M is a 0/1 matrix with a single 1 in each column.  Randomized computation is similar.  This time the state is a vector of nonnegative real numbers summing to one, and $x\mapsto Mx$ is a valid transition for any $M$ with nonnegative entries whose columns each sum to one.  In each case, this picture neglects the fact that some transformations of $n$ bits are easier than others.  To represent a single operation on, say, two bits, we use a matrix $M$ for which, again, $M_{i,j}=0$ whenever $i$ and $j$ disagree on some bit not being acted upon, and $M_{i,j}$ depends only on the values of $i,j$ for the bits being acted upon.  This represents the idea that if we don't act on a bit, it shouldn't change or affect what happens.

Thus the key differences between randomized and quantum computation appear to be ``merely'' the shift from real to complex numbers (and even allowing negative reals captures most of this difference) and the shift from $\ell_1$ to $\ell_2$ norm for the state; that is, quantum states have the squared absolute values of their amplitudes summing to one, while probabilities sum to one without squaring.  However, the fact that different branches of a computation can have different phases means that when they recombine their amplitudes can either add up (called {\em constructive interference}) or can cancel each other (called {\em destructive interference}).  If ``0'' is represented by the vector $\bpm 1 \\ 0 \epm$ and ``1'' is represented by $\bpm 0 \\ 1 \epm$, then the NOT operation can be expressed (quantumly or classically) as $\bpm 0 & 1 \\ 1 & 0 \epm$.  Geometrically, this is a rotation by $\pi/2$.
However, only a quantum computer can perform the following ``square root of NOT,'' which corresponds to a rotation by $\pi/4$.
$$\sqrt{\text{NOT}} = \bpm 1/\sqrt{2} &  - 1/\sqrt{2} \\
1/\sqrt{2} &  1/\sqrt{2}  \epm.$$
 
If we start with the ``0'' state and apply $\sqrt{\text{NOT}}$, then we obtain the state $\bpm 1/\sqrt 2 \\ 1/\sqrt 2 \epm$.  Were we to measure, the outcomes 0 or 1 would each occur with probability 1/2.  However, if we apply $\sqrt{\text{NOT}}$ a second time before measuring, then we always obtain the outcome 1.  This demonstrates a key difference between quantum superpositions and random mixtures; placing a state into a superposition can be done without any irreversible loss of information.
When we have $n$ qubits, superposition and interference enable us to achieve greater computational advantages.  One famous example is Grover's algorithm, which, given a binary function $f$ on $n$ bits, allows us to search for an input $x\in \{0,1\}^n$ with $f(x)=1$ using $\approx 2^{n/2}$ evaluations of $f$; a square-root advantage.  What makes Grover's algorithm possible is the fact that probabilities are the squares of amplitudes.  Thus, a uniform superposition over $2^n$ states assigns amplitude $1/\sqrt{2^n}$ to each one.  Moreover, one can show that with effort comparable to that required to calculate f, it is possible to increase the amplitude of each target $x$ (i.e. with $f(x)=1$) by roughly $1/\sqrt{2^n}$ .  This breaks down only when their total amplitude is large.  Thus, the total effort is on the order of $2^{n/2}$, or more generally $\sqrt{2^n / M}$ if there are $M$ solutions.
 
A more dramatic speedup over classical computing occurs with Shor's algorithm for factoring integers.  Shor's algorithm has a substantial classical component which reduces the problem of factoring to the more abstract problem of period finding.  This problem takes as input a function f on the integers $\{0, 1,…, 2^n-1\}$ with the property that $f(x)=f(y)$ if and only if $x-y$ s divisible by $r$ for some hidden period $r$.  The goal is to find $r$.  Since $r$ can be exponentially large in $n$, classical computers require exponential time to find it if they are required to treat $f$ as a black box.  

However, it is possible for a quantum computer to create a superposition of states with amplitude $\approx \sqrt{r/2^n}$ in each of $z, z+r, z+2r, \ldots$  for some randomly chosen $z\in \{0,1,\ldots,r-1\}$.  So far this is analogous to what a probabilistic computer would do if it chose a random $x$, calculated $f(x)$, and considered the distribution on $x$ conditioned on this choice of $f(x)$.  The next step is uniquely quantum, and involves applying a unitary matrix called a quantum Fourier transform (QFT).  This matrix has $y,z$ entry equal to $e^{2\pi i y z/2^n}/\sqrt{2^n}$.  Performing it efficiently is possible by adapting the classical fast Fourier transform (FFT) to the quantum setting, but its action is dramatically different, since it transforms the amplitudes of a quantum state rather than transforming a list of numbers as the classical FFT does.  Applying the QFT in this case maps our superposition to a state where the amplitude of $y$ is $\approx \frac{r}{2^n}(1 + e^{\pi i \frac{yr}{2^n}} + e^{\pi i \frac{2yr}{2^n}} + e^{\pi i \frac{3yr}{2^n}} + \ldots)$.  If $yr$ is (approximately) divisible by $2^n$ then this sum will be large, and if $yr$ is far from divisible by $2^n$ then it will involve many complex numbers with different phases that tend to cancel out (as can be verified by a quick calculation).  Thus, measuring $y$ will return an answer that is close to a multiple of $2^n/r$.  Finally, a classical continued fraction expansion can recover $r$.

While quantum simulation is likely to become more practically important once we develop new public-key cryptosystems that resist quantum attacks, Shor's algorithm is a far more surprising demonstration of the power of quantum computing because the problem it solves has no obvious relation to quantum mechanics.  After Shor's algorithm, it also became increasingly difficult to believe that quantum mechanics might be efficiently simulatable classically.  Previously, many scientists had hoped for a model of quantum mechanics that would tame some of its most counterintuitive features, such as exponentially large state spaces, while remaining consistent with our observations.  But now we know that these models cannot be much simpler than quantum mechanics itself (or at least no easier to simulate) unless factoring and period-finding can be performed much more quickly on classical computers.  This situation is like the development of NP-completeness, which showed that efforts in disparate areas to solve NP-complete problems were in a sense equivalent.

Shor's algorithm and Grover's algorithm are the most famous quantum algorithms, but are not the only two.  One recent algorithm~\cite{HHL08} solves large linear systems of equations; that is, given a matrix $A$ and a vector $b$, it finds $x$ such that $Ax=b$.  However, unlike classical algorithms for the problem, in the quantum version $x$ and $b$ are quantum states (and thus can be $2^n$-dimensional vectors using only $n$ qubits).  Further, $A$ needs to be sparse and given by an implicit representation so that given any row index $i$ it is possible to efficiently find all the nonzero $A_{ij}$.  These restrictions make it possible in principle to solve exponentially large systems of equations in polynomial time.  However, there is an important caveat: the runtime of the algorithm also scales with the condition number of $A$, a parameter which is related to the numerical instability of solving the system of equations classically.  Finding a natural scenario in which this algorithm can be used remains a compelling open problem.

Another recent algorithmic development is a quantum analogue of the Metropolis sampling algorithm~\cite{qmetropolis11}.  Classically, the Metropolis method is a technique for sampling from hard-to-analyze distributions over exponentially large state spaces (indeed, exponential speedup via use of randomness is a less glamorous example of how powerful a change of computational model can be).  Its amazing range of applications includes statistical inference and approximation algorithms for the permanent of a nonnegative matrix, but it was originally developed to sample from the thermal distribution of a system.  If state $x$ has energy $E(x)$, then the thermal distribution at temperature $T$ assigns $x$ probability proportional to $e^{-E(x)/T}$, so that lower temperatures push the system harder into low-energy configurations.  The quantum Metropolis algorithm, by analogy, produces quantum states in thermal distributions.  Like its classical cousin, the quantum Metropolis algorithm takes longer to produce lower-temperature states, as this corresponds to solving harder optimization problems, but proving rigorous bounds on its runtime is difficult in all but a handful of cases.  Without formal proofs, we can nevertheless run the classical Metropolis algorithm and observe empirically that it works quickly in many cases of interest.  An empirical evaluation of the quantum Metropolis will of course have to wait for the arrival of a large-scale quantum computer.  But we may already know enough tools, if we can figure out how to combine them, to develop new quantum algorithms that use quantum Metropolis as a subroutine, just as the classical algorithm for the permanent used classical Metropolis.

	New proposed uses for quantum computers continue to arise.  One exciting development is the increasing number of applications being found for quantum computers with 10 or so qubits; small enough to be easily simulatable classically, while large enough to interact in ways not previously demonstrated.  Such quantum computing devices could be used to improve precision quantum measurements (e.g. in atomic clocks, or to detect gravity waves), as ``quantum repeaters'' in a network to distribute entanglement for use in cryptographic protocols or even to construct arrays of telescopes that could synthesize apertures of unlimited size.  Just as the usage model of classical computers has gone beyond the Turing machine model, quantum computing devices are likely to be more flexible than we can currently imagine.

\section{Science through the (quantum) algorithmic lens}
Building large-scale quantum computers will undoubtedly cause large and unpredictable changes in how we think about science.  But even without a physical quantum computer, our theoretical progress so far has led to many conceptual advances.

One major advance is the idea of separating the informational content of quantum mechanics from the physics.  Teaching these together, as is done now, mixes counterintuitive features such as measurement and entanglement with the mathematically complicated picture of the Schr'\"{o}dinger equation as a PDE, and of states as living in the infinite-dimensional space of functions on $\bbR^3$.  It is as though probability were only taught in the context of statistical mechanics, and the first distributions that students saw were the thermal distributions of ideal gases.  Instead, I believe quantum mechanics would make more sense if its implications for information were taught first, and then this framework was used to explain atoms, photons, and other physical phenomena.

Starting students are not the only practitioners of quantum mechanics who can benefit from the quantum informational perspective.  Many phenomena involving quantum mechanics are related to issues such as entropy, entanglement or correlation that have physical relevance but are best described in terms of information.  One early success story concerns the problem of finding the lowest-energy state of a quantum system (of $n$ qubits, with pairwise interactions).  For classical systems, such problems are NP-complete, except in special cases, such as when the systems are arranged in a line or on a tree.  For quantum systems, the energy-minimization problem is QMA-complete.  QMA stands for ``Quantum Merlin-Arthur games,'' and is believed to be (roughly) as far out of the reach of quantum computers as NP is out of reach of classical computers.

This gives some theoretical justification to the empirically observed phenomenon that many physical systems, such as glasses, have trouble finding their ground states.  Surprisingly, energy minimization is QMA-complete even for systems on a line with identical nearest-neighbor interactions, contrary to the previously held intuition of physicists that the 1-D case should be easy.

In other cases, the quantum information perspective yields positive results, and even new classical algorithms.  One example relates to the gap between the lowest energy level of a system and the second-lowest level.  Physically, this energy gap corresponds to the mass of an excitation.  Some excitations, like photons or the vibrations of a solid, will be massless while others, like extra electrons moving through semiconductors, will have effective masses.  From this picture, we expect that a small gap will correspond to long-range correlations, while a large gap will mean that correlations rapidly decay.  The reality turns out to be more subtle.  For any single observable, say the strength of the magnetic field at a particular point, correlations indeed decay rapidly when the gap is large.  But the state of the system as a whole may still exhibit long-range correlations. A good analogy for the situation arises from considering fixing a thousand random one-to-one functions on the space of a million bits.  If the function f is chosen randomly, then the pair $(x,f(x))$ will have a nearly-maximal amount of mutual information, meaning that learning $x$ will narrow down the number of possibilities for $f(x)$ from $2^{1,000,000}$ down to 1000.  On the other hand, any individual bit in $x$ will be nearly uncorrelated with any particular bit in $f(x)$.  This argument also works if the random functions are replaced by an expander graph with suitable parameters, meaning that this behavior can also be achieved constructively.  Guided by this intuition, researchers have developed quantum analogues of expander graphs that demonstrate systems with a large gap and rapidly-decaying correlations of any individual observable, but nevertheless a large amount of overall mutual information between distant pieces of the system.
Why should we care about these different kinds of correlations, apart from the desire to use theory to predict observable quantities in the real world?  One exciting application is the old problem of simulating quantum systems on classical computers.  If a system of $n$ qubits had no entanglement, then the number of parameters required to describe it would not be $2^n$, but merely $2n$.  If the qubits are arranged on a line, with correlations roughly bounded in range to some short distance $k$, then the number of parameters scales as $n\cdot \exp(k)$.  Thus, controlling correlations in quantum systems also helps us simulate them efficiently.

This line of research can be seen as part of a larger project to divide quantum systems into those that are efficiently simulatable classically (say because they have limited entanglement), and those that are capable of universal quantum computing.  On the other hand, a handful of systems of apparently intermediate complexity have been found, which are neither known to be classically simulatable, nor universal for quantum computing.  These include systems of non-interacting photons, as well as the case when the noise rate is too high for known quantum error-correcting codes to function, but too low to rule out the possibility of large-scale entanglement.  Resolving the complexity of these boundary cases is a fascinating source of open problems.

	The scientific benefits of the quantum information perspective are not restricted to quantum mechanics.  Some important problems with no apparent relation to quantum mechanics relate to performing linear algebra on multidimensional arrays.  For example, given a 3-d collection of numbers $A_{ijk}$, with $i,j,k$ ranging over $\{1,\ldots,n\}$, how hard is it to compute the following analogue of the largest singular value: $\max \sum_{i,j,k} A_{i,j,k} x_i y_j z_k$ over all unit vectors $x,y,z$?  While computing this to accuracy $1/\poly(n)$ can be readily shown to be NP-hard, for the often more realistic case of requiring constant accuracy, the only hardness result known involves quantum techniques, as does the most promising classical algorithm.   One possible reason for the effectiveness of the quantum information perspective here is that multidimensional arrays correspond naturally to entangled states, and our increasingly quantitative understanding of entanglement often yields results about linear algebra that are more widely useful.  Linear algebra appears to be gaining in importance within theoretical computer science.  Examples include using Fourier analysis on the Boolean cube and taking advantage of the way that graphs and matrices can be viewed interchangeably, thereby mixing combinatorial and algebraic pictures.  In the future, I expect that our view of linear algebra and probability will become increasingly shaped by tools from quantum information.

	Finally, most of the contributions of quantum information to computer science are at this stage theoretical, since large-scale quantum computers have yet to be built.  But once they are, we can expect them to be used in ways that theorists will struggle to explain, just as we see classically with successful heuristics such as the simplex algorithm.  For example, periodic structure was exploited by Shor's algorithm, and can be applied to obtain several other exponential speedups.  In the future, we might use tools such as period-finding for exploratory data analysis, much as linear regression is used today.  Many of these new tools will initially be deeply counter-intuitive, but as we master them, they promise radically new ways of looking at the world scientifically.

\section*{Further reading}
\begin{itemize}
\item M. A. Nielsen and I. L. Chuang. {\em Quantum Computation and Quantum Information}. Cambridge University Press, 2001.

\item J. Preskill. Lecture notes for Ph219.  \url{http://theory.caltech.edu/~preskill/ph229/}

\item D. Mermin. Lecture notes for CS483. \url{http://people.ccmr.cornell.edu/~mermin/qcomp/CS483.html}
\end{itemize}

\end{document}